\documentclass[sigconf,nonacm]{acmart}
\setcopyright{none} 
\AtBeginDocument{%
  \providecommand\BibTeX{{%
    \normalfont B\kern-0.5em{\scshape i\kern-0.25em b}\kern-0.8em\TeX}}}

\acmISBN{978-1-4503-XXXX-X/18/06}

\begin{document}
\cfoot{\thepage}
\title{The Friendship Paradox and Social Network Participation}

%%todo: add for actual publication
\author{Ahmed Medhat}
\affiliation{%
\institution{Meta Core Data Science}
\country{Menlo Park, CA, USA}
}
\email{amedhat@meta.com}

\author{Shankar Iyer}
\affiliation{%
\institution{Meta Core Data Science}
\country{Menlo Park, CA, USA}
}
\email{shankar94@meta.com}

\begin{abstract}
The friendship paradox implies that a person will, on average, have fewer friends than their friends do. 
Prior work has shown how the friendship paradox can lead to perception biases regarding behaviors that correlate with the number of friends: for example, people tend to perceive their friends as being more socially engaged than they are. 
Here, we investigate the consequences of this type of social comparison in the conceptual setting of content creation (“sharing”) in an online social network. 
Suppose people compare the amount of feedback that their content receives to the amount of feedback that their friends' content receives, and suppose they modify their sharing behavior as a result of that comparison. 
How does that impact overall sharing on the social network over time? 
We run simulations over model-generated synthetic networks, assuming initially uniform sharing and feedback rates. 
Thus, people’s initial modifications of their sharing behavior in response to social comparisons are entirely driven by the friendship paradox. 
These modifications induce inhomogeneities in sharing rates that can further alter perception biases. 
If people's responses to social comparisons are monotonic (i.e., the larger the disparity, the larger the modification in sharing behavior), our simulations suggest that overall sharing in the network gradually declines. 
Meanwhile, convex responses can sustain or grow overall sharing in the network. 
We focus entirely on synthetic graphs in the present work and have not yet extended our simulations to real-world network topologies. 
Nevertheless, we do discuss practical implications, such as how interventions can be tailored to sustain long-term sharing, even in the presence of adverse social-comparison effects.
\end{abstract}


\ccsdesc[500]{Applied Computing~Sociology}
\ccsdesc[500]{Networks~Online Social Networks}

\keywords{social networks, graph theory, friendship paradox, social comparison}

\maketitle

\section{Introduction}

In a network that represents friendship relations between a group of people, if we choose a person and then choose one of that person's friends, the friend so chosen will be relatively likely to be someone who has many friends.
This oversampling is the origin of the “friendship paradox”, the phenomenon that individuals in networks will often find that their friends have (on average) more friends than they do.
In 1991, Feld demonstrated that a particular version of the paradox will occur in any network with non-zero variance in the degree distribution \cite{feld1991your}.
The paradox has since been observed in both social and non-social contexts \cite{pires2017friendship}, with implications in both offline \cite{nettasinghe2019diffusion, nettasinghe2019your, geng2019sentinel} and online networks \cite{hodas2013friendship,higham2019centrality,yan2022does}. 

More recent work has generalized the paradox to other network properties and behaviors \cite{hodas2013friendship, kooti2014network, eom2014generalized, higham2019centrality, bollen2017happiness, lee2019impact}.
Eom and Jo showed that a “paradox” occurs for academic productivity in scientific-collaboration networks: a researcher's collaborators are, on average, more productive than them \cite{eom2014generalized}, and Hodas et al. showed how an individual's friends are on average more active and have access to more information\cite{hodas2013friendship}. Kooti et al. showed that these generalized paradoxes can originate in both degree-attribute correlations and assortativity of links with respect to the attribute value \cite{kooti2014network}.

Researchers have also shown how friendship-paradox-based perception biases can impact social comparison and behavior in networks. For example, while Scissors et al. observed a “like paradox” on Facebook, they found evidence that people care more about who likes their content, than they do about like volume\cite{scissors2016s}, while Bollen et al. observed a "happiness paradox" as a correlate of the friendship paradox \cite{bollen2017happiness}. Further, Jackson showed that in a network with incomplete information, the friendship paradox induced perception biases lead to amplifying average engagement in the network \cite{jackson2019friendship}

Following these lines of research, this paper investigates how the interplay of the friendship paradox and social comparison affects content contribution in online social networks. 
We specifically consider an online social network where contributions are “posts” or “shares” and where social comparison originates from people receiving feedback and seeing their friends receive feedback.
Through running simulations over model-generated synthetic networks, we study how, in the absence of any baseline sharing rate differences, the local friendship paradox and people’s response to feedback-related social comparisons combine to shape individual and overall contribution rates in a network. 

\section{Model Formulation}

We propose a simplified model of sharing in an online social network, in which an individual's perceived disparity between their feedback and their friends' feedback depends upon three factors:
\begin{enumerate}
    \item the local structural friendship paradox, where a person's friends can can have more friends on average than they do.
    \item sharing bias, where sharing amongst a person's friends may skew towards friends with greater or lower degrees.
    \item engagement bias, where a person's friends may receive more feedback per friend than they do.
\end{enumerate}
We will demonstrate below, how the local paradox alone is sufficient to establish a baseline feedback disparity in absence of a sharing or engagement bias. This then alters sharing bias, creating a continuous feedback loop between sharing bias and feedback disparity.

To frame this mathematically, nodes of the network represent participants in the social network, and links of the network represent friendship ties between participants. The adjacency matrix $A_{u, v} = 1$ if $u$ and $v$ are friends and 0 otherwise. 
With each person having a degree $d_u$, each person in the network experiences a local friendship paradox $lp_u$:
\begin{equation}
lp_u = \frac{\frac{\sum_v A_{u, v}d_v}{\sum_v A_{u, v}}}{d_u}
\end{equation}
where $\frac{\sum_v A_{u, v}d_v}{\sum_v A_{u, v}}$ is the \textbf{average friend degree} \\

In a sequence of time steps $t = 0, 1, 2,\ldots$, each participant $u$ shares content at a rate $r_u(t)$.
This can be interpreted as the number of pieces of content that $u$ shares per week, with each time step representing some longer period of time (e.g., a month, a year, etc.). 
All rates are initialized to $r_u(0) = 1$. 
This means that there is no initial \textbf {sharing bias} in our model, although sharing bias can emerge over time through the mechanisms that we describe below. 
We can quantify the sharing bias $sb_u(t)$ as follows:
\begin{equation}
sb_u(t) = \frac{\frac{\sum_v A_{u, v} r_v(t) d_v}{\sum_v A_{u, v} r_v(t)}}{\frac{\sum_v A_{u, v}d_v}{\sum_v A_{u, v}}}
\end{equation}

The product of local paradox $lp_u$ and sharing bias $sb_u(t)$ can be thought of as the \textbf{effective or weighted local paradox} $wlp_u(t)$, since it represents the average degree of friends to whom $u$ has been exposed through shared content, divided by $u$'s degree.

\begin{equation}
wlp_u(t)=lp_u sb_u(t) = \frac{\frac{\sum_v A_{u, v}d_v}{\sum_v A_{u, v}}}{d_u}\frac{\frac{\sum_v A_{u, v} r_v(t) d_v}{\sum_v A_{u, v} r_v(t)}}{\frac{\sum_v A_{u, v}d_v}{\sum_v A_{u, v}}}=\frac{\frac{\sum_v A_{u, v} r_v(t) d_v}{\sum_v A_{u, v} r_v(t)}}{d_u}
\end{equation}

Further, engagement bias can modulate the effect of the weighted local paradox if an individual's friends' content receives feedback at a different rate per friend. 
We define $e_v(t)$ as the \emph{feedback per friend} that each person $v$ in the network receives.
This means that, when sharing rate $r_u(t)>0$ at time t, $u$ receives $f_u(t) = e_u(t)d_u$ feebdack per piece of content, while seeing their friends receive: 
\begin{equation} 
f^{nbr}_u(t) = \frac{\sum_v A_{u, v} r_v(t) e_v(t)d_v}{\sum_v A_{u, v} r_v(t)} 
\end{equation}
Then, $z_u(t) = \frac{f^{nbr}_u(t)}{f_u(t)}$ is the ratio of these two feedback levels.
For simplicity, we assume that that there is no \textbf {engagement bias} in our model; the amount of feedback per friend that each piece of content receives is identical, with $e_v(t) = K$. 
The derivation below shows how, under this assumption, perceived feedback disparity $z_u(t)$, for a person u with $r_u(t)>0$, is equivalent to the weighted local paradox:

\begin{equation} 
f^{nbr}_u(t) = \frac{\sum_v A_{u, v} r_v(t) K d_v}{\sum_v A_{u, v} r_v(t)} = K \frac{\sum_v A_{u, v} r_v(t) d_v}{\sum_v A_{u, v} r_v(t)}
\end{equation}

\begin{equation}
z_u(t) = \frac{f^{nbr}_u(t)}{f_u(t)} = \frac{K \frac{\sum_v A_{u, v} r_v(t) d_v}{\sum_v A_{u, v} r_v(t)}}{K d_u} = wlp_u(t) 
\end{equation}

Therefore, to calculate feedback disparity at every step for a person who is producing content, we simply compute their weighted local paradox at that step.

With this formula in mind, the participant then updates their sharing rate in response to the social comparison implied by feedback disparity $z_u(t)$:
\begin{equation} r_u(t + 1) = \begin{array}{cc}\{ & \begin{array}{cc} 0, & r_u(t)= 0 \\ r_u(0),& r_u(t) > 0 , \sum_v A_{u, v} r_v(t) = 0 \\ drf(z_u(t)) r_u(0),& r_u(t) > 0 , \sum_v A_{u, v} r_v(t)) > 0  \\ \end{array} \end{array}
\end{equation}
Outcome 1 means that, if a person’s sharing rate hits 0, it will stay at 0. 
Outcome 2 means that, if all of a sharer's friends have stopped sharing, the sharer will continue sharing at their baseline rate. 
Outcome 3 means that, if the prior two conditions do not hold, a person will modify their sharing rate by multiplying their baseline sharing rate $r_u(0)$ by the disparity response function output. 

We refer to \begin{math}drf(x)\end{math} as the feedback \textbf{disparity response function}, as it controls how participants update their sharing rate in response to disparities between feedback on their content and that of their friends. 
Because we set $r_u(0) = 1$, these disparities are initially determined by the local structural friendship paradox that each participant \begin{math}u\end{math} experiences. 
As $t$ progresses, the disparities can be driven by a combination of the structural friendship paradox and inhomogeneous sharing patterns amongst \begin{math}u\end{math}’s friends. 
Each participant \begin{math}u\end{math} then experiences a feedback disparity \begin{math} z_u(t) \end{math} that determines future updates to their sharing rate. 
Since each step in our model is deterministic, the sharing trajectory is uniquely determined by the structure of the network and the choice of \begin{math}drf(x)\end{math}.

Hereafter, we abbreviate our model as \textbf {FIT}, short for Friendship Paradox Induced Sharing Trajectories.

\subsection{Special Case of a Disparity Threshold}

We explore several network topologies and disparity response functions in this work. 
However, one particular choice of disparity response function merits special mention: the case where the \begin{math}drf(x)\end{math} is a negative step function, equal to 1 when \begin{math}z_u(t)\end{math} is less than or equal to \begin{math}z*\end{math} and 0 otherwise. 
Here, we refer to the value \begin{math}z*\end{math} as the \textbf{disparity threshold}. 
If $u$ experiences a feedback disparity \begin{math} z_u(t) \end{math} that is greater than this threshold, $u$ permanently stops sharing content. 
This particularly severe choice of a drf leads simulations to quickly converge, allowing for exploration of asymptotic sharing states. 
It also offers a particularly simple setting for the introduction of additional mechanisms into our model, such as an \textbf{activity threshold} a*. 
As in threshold models of collective behavior, the threshold a* is the fraction of an individual $u$'s neighbors that must be sharing in time step $t$ for $u$ to continue sharing thereafter \cite{macy2020threshold}. 

Why is this an interesting enrichment of the model? 
The churn of sharers due to feedback disparities tends to exacerbate the feedback disparities that remaining sharers experience. 
Since, with a disparity threshold, the sharers who are most likely to continue sharing indefinitely are those whose friends have lower degrees, and therefore receive less feedback, than they do. 
If $a* > 0$ however, these same individuals also become vulnerable to churn, once enough of their lowest-degree friends stop sharing. 
The churn of these high-degree sharers can, in turn, produce more favorable local disparities for their friends who are still sharing. 
Thus, in the enriched model, there can conceivably be stabilized low-disparity clusters of participants that remain sharing indefinitely. 
This is a possibility that we explore through simulations below.

\subsection{Florentine Families Network }

We will now illustrate our model on a commonly studied small-scale network: the “Florentine Families Network” (FFN). 
In this network, nodes represent prominent families in 15th century Florence and links represent business or marital ties between them \cite{breiger1986cumulated}. 
Although the structural properties of FFN are distinct from online social networks, it is commonly used to get a qualitative understanding of the behavior of various centrality measures. 
Along the same lines, it can be used to gain a qualitative understanding of who shares over time in our model.

\begin{figure}[h]
  \centering
  \includegraphics[width=\linewidth]{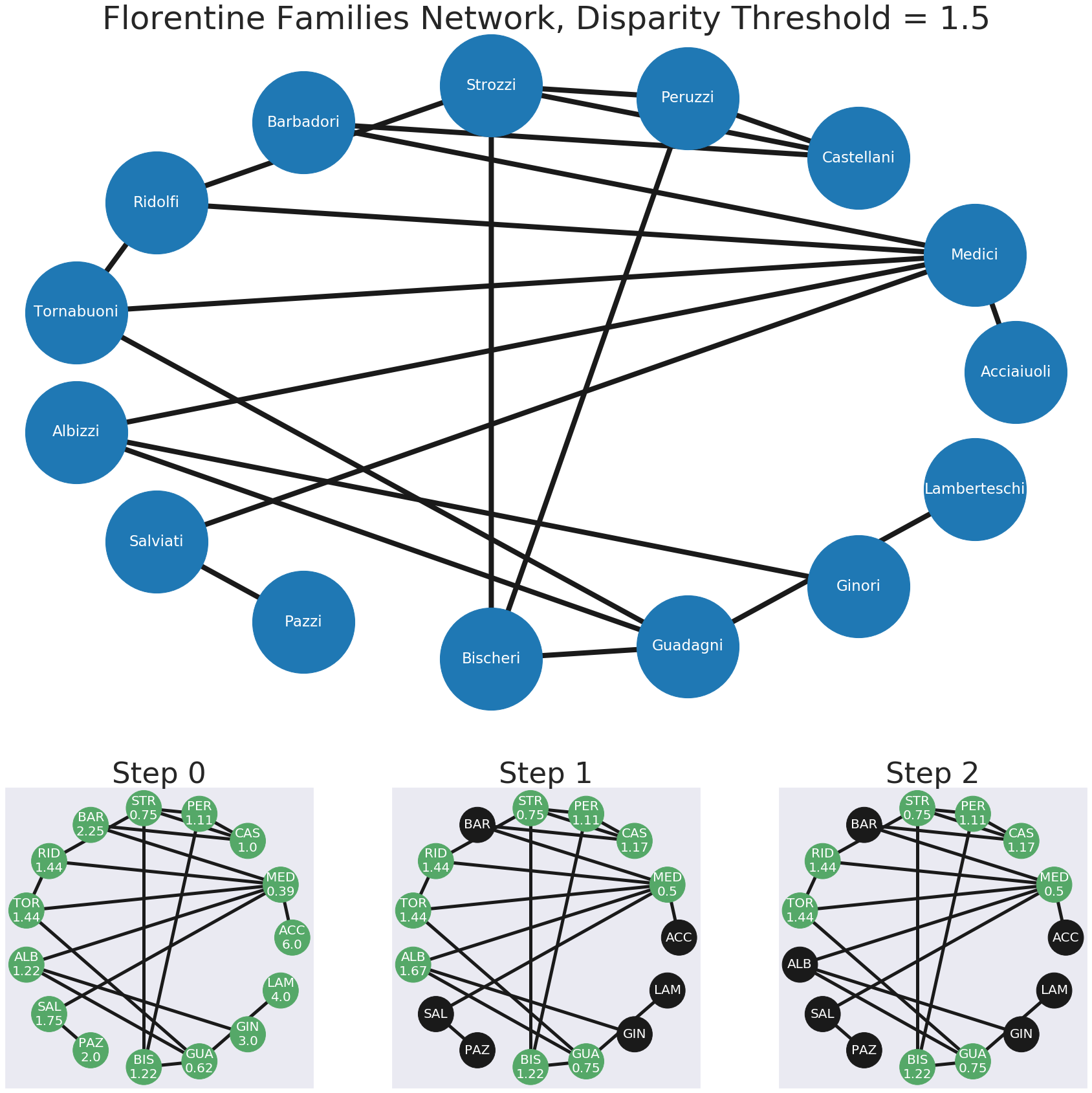}
  \caption{Illustrating the FIT model using the Florentine Families Network}
  \label{florentine_simulation}
\end{figure}

In the diagrams in figure \ref{florentine_simulation}, we show results of running the FIT simulation on the FFN with $z* = 1.5$. 
Meaning that people whose friends receive more than 1.5 times the feedback that they receive will stop sharing in the next step. 
Green nodes represent those who are sharing, and black nodes represent those that have stopped sharing. 
Numerical values inside each green node signify the feedback disparity value at that step. 
The approach converges after 2 steps, with 6 nodes ceasing to share by the step 1 and one more need ceasing to share by the step 2.

\section{Simulation Setup}

We run the FIT model on synthetically generated Erdős–Rényi \cite{erdHos1960evolution} and Barabási-Albert networks \cite{barabasi1999emergence}. 
Power-law degree distributions, such as those observed in Barabási-Albert networks, amplify local paradoxes relative to random networks.
Thus, we simulate on these two classes of networks in order to investigate the impact of the stronger paradoxes in the Barabási-Albert case.
In future work, we plan to explore our model as applied to real-world networks. 

\begin{figure}[h]
  \centering
  \includegraphics[width=\linewidth]{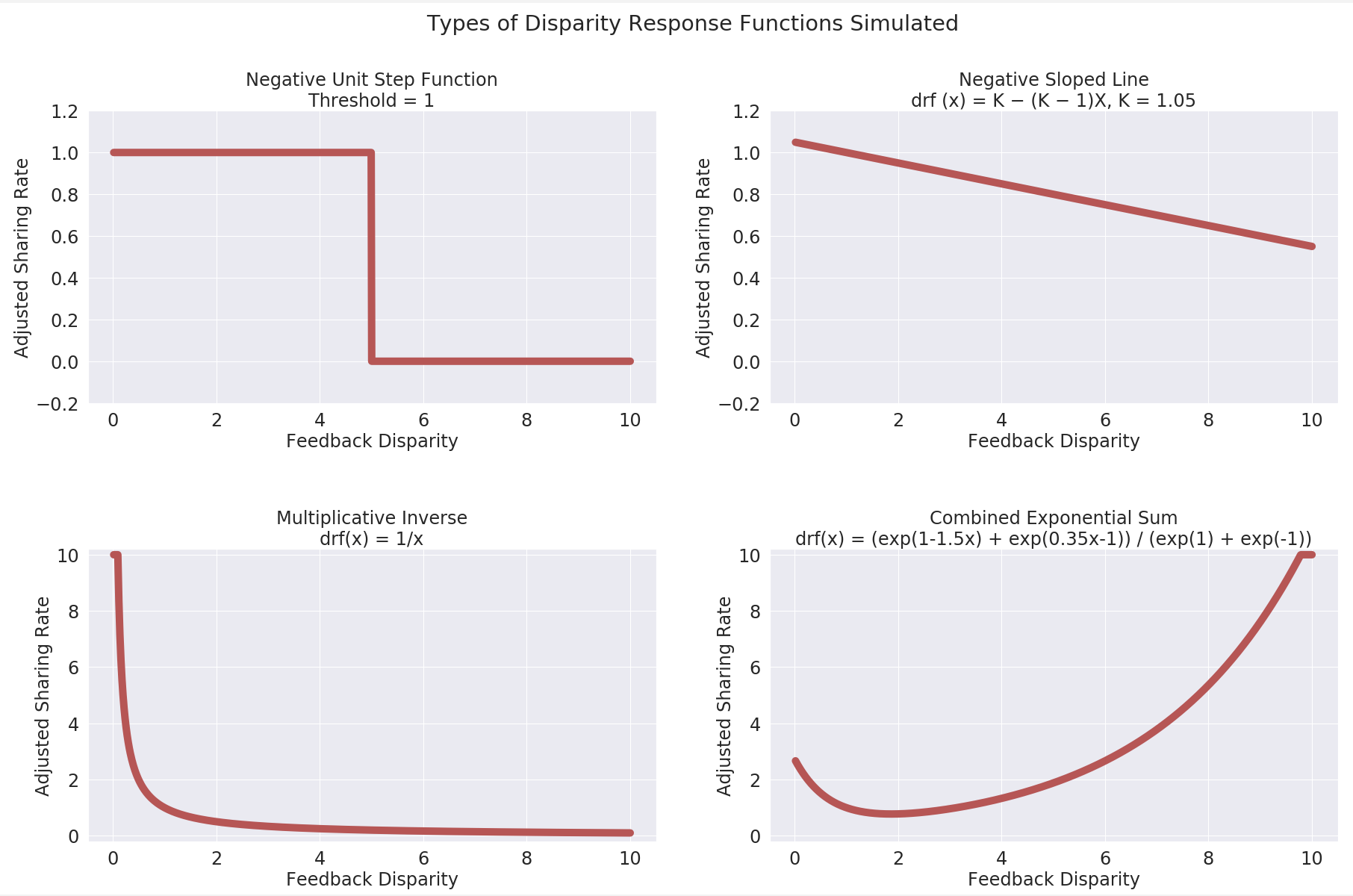}
  \caption{Types of Disparity Response Functions Simulated}
  \label{drfs}
\end{figure}

We set up our simulations as follows:
\begin{itemize}
    \item We generate 10 Erdős–Rényi and 10 Barabási-Albert networks, each with 3000 nodes and network densities going from 0.7\% to 7\% (average degrees going from 20 to 200)
    \item We generate a series of different disparity response functions (DRFs), as shown in figure \ref{drfs}. All functions guarantee that for a feedback disparity of 1, sharing rate adjusts to 1. We model 3 monotonically decreasing functions relative to feedback disparity, and a convex function;
    \begin{enumerate}
    \item A negative unit step function, where sharing rate adjusts to 0 if threshold k is exceeded. We generate 21 different function instances with \begin{math}K\end{math} taking on values from 0.5 to 2.6.
    \begin{equation} drf(x)=\begin{array}{cc}\{ & \begin{array}{cc} 1, & x\leq K \\ 0, & x>K \\ \end{array} \end{array} \end{equation}
    \item A negative sloped linear function, with intercept \begin{math}K\end{math}. People take on sharing rates going from \begin{math}K\end{math} for a feedback disparity of 0 to a sharing rate of 0 for a disparity of \begin{math}K/(K-1)\end{math}. We generate 3 different function instances, with the \begin{math}K\end{math} intercept taking on values from 1.05 to 2.05
    \begin{equation}drf(x)=K-(K-1)x\end{equation}
    \item A multiplicative inverse function.
    \begin{equation}drf(x) = \frac{1}{x}\end{equation}
    \item A convex exponential sum function, where people experience a drop in sharing rate as feedback disparity goes from 0 to ~2, then start to experience an increase in sharing rate when disparity becomes greater than 2, and we cap that sharing rate at 10 per step.
    \begin{equation}drf(x) = min(\frac{e^{1-1.5X} + e^{0.35X - 1}}{e^{1} + e^{-1}}, 10)\label{convex_drf_equation}\end{equation}
    \end{enumerate}
\end{itemize}

Each simulation follows these steps:

\begin{enumerate}
    \item Initiate each person in the network with $r_u(0) = 1$. We assume no engagement bias in terms of engagement per friend per post. This leads to the initial feedback disparity being equal to the local paradox.
    \item At each step $t+1$,
    \begin{enumerate}
        \item For all people, we use the current feedback disparity value, \begin{math}z_u(t)\end{math} as input to the disparity response function, \begin{math}drf(z_u(t))\end{math} to adjust each person’s sharing rate, \begin{math}r_u(t+1)\end{math}, as detailed in the model formulation section.
        \item Using the network’s revised sharing rates, we adjust feedback disparity scores of all people, \begin{math}z_u(t+1)\end{math}.
    \end{enumerate}
\item We run the simulation until it converges to no people altering their sharing rates, or until the simulation runs for 52 steps. If each step signifies sharing rates over 1 week, 52 steps capture sharing trajectories over a year’s time.

\end{enumerate}

\section{Results}
\subsection{Negative Unit Step Function Simulations}

\begin{figure}[h!]
  \centering
  \includegraphics[width=\linewidth]{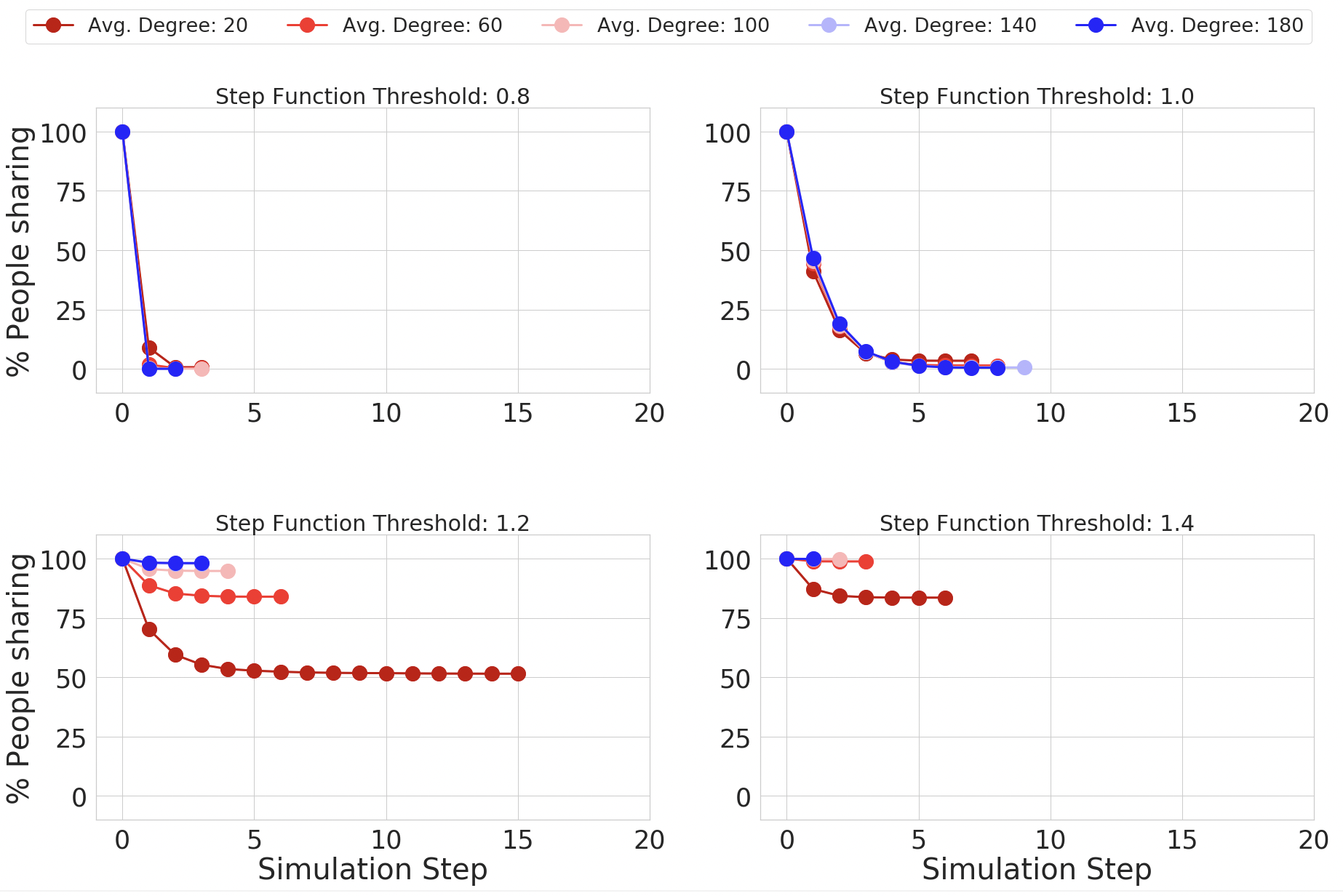}
  \caption{Negative Unit Step Function Simulations run over 5 Erdős–Rényi Networks of different average degrees}
  \label{negative_step_erdos}
\end{figure}

Figure \ref{negative_step_erdos} reports results for negative unit step function disparity response functions on Erdős–Rényi networks. We observe that:

\begin{enumerate}
    \item Almost all simulations show a gradual monotonically decreasing sharing rate trending towards an equilibrium point.
    \item A threshold of 1, where people stop sharing if they experience any feedback disparity, leads to almost no one sharing in the network for all 10 simulated Erdős–Rényi Networks.
    \item Simulations with thresholds greater than 1 have meaningfully higher sharing rate points of equilibrium.
    \item The larger the average degree in the network, the higher this equilibrium point is for thresholds greater than 1.
\end{enumerate}

\begin{figure}[h]
  \centering
  \includegraphics[width=\linewidth]{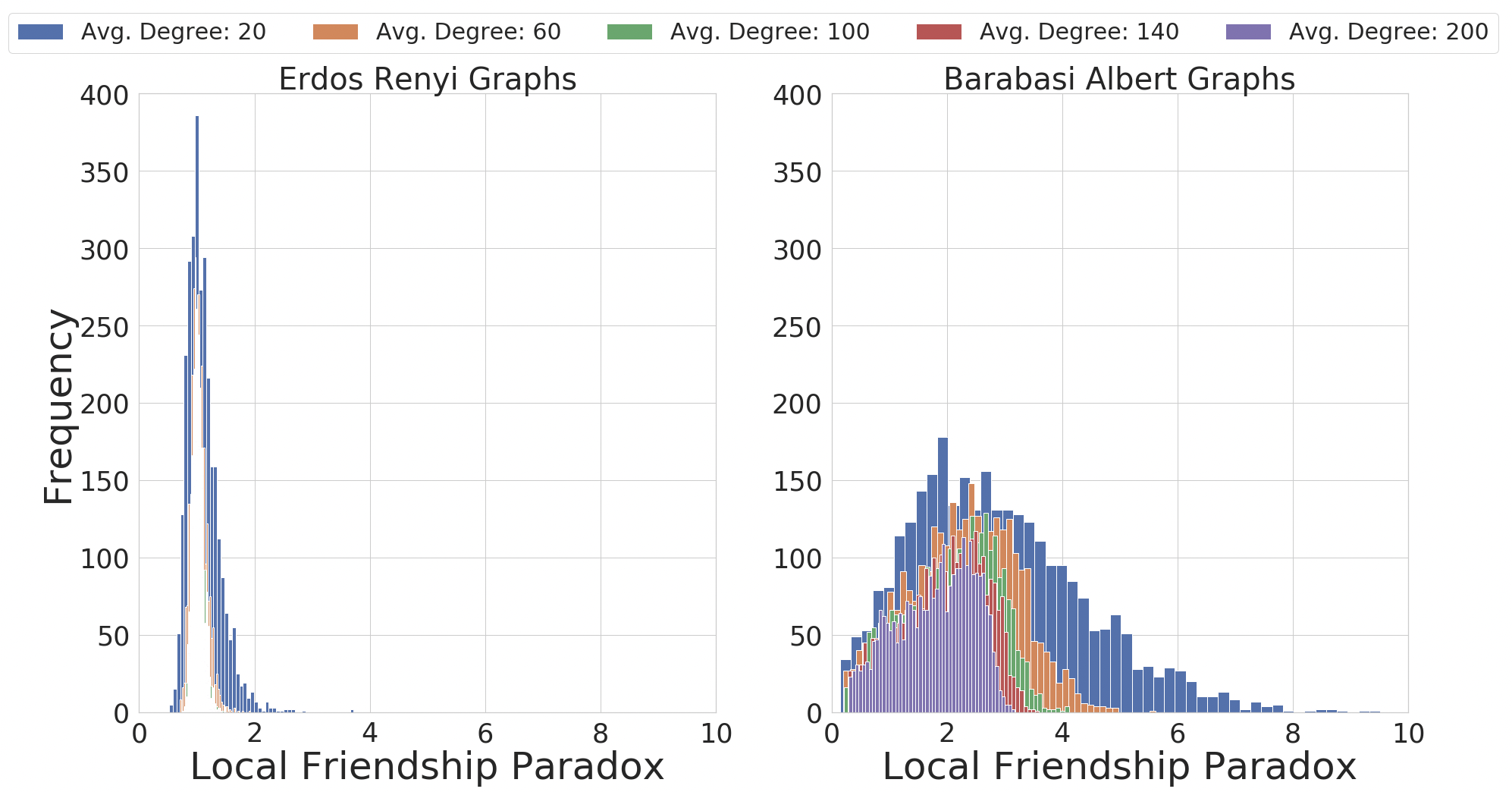}
  \caption{Local Friendship Paradox Distributions for Simulated Networks of different average degrees}
  \label{paradox_distributions}
\end{figure}

As shown in figure \ref{paradox_distributions}, point 3 is explained by the narrow distribution of local paradoxes in Erdős–Rényi networks and point 4 is explained by that distribution being wider the smaller the average degree is in the network.

\begin{figure}[h]
  \centering
  \includegraphics[width=\linewidth]{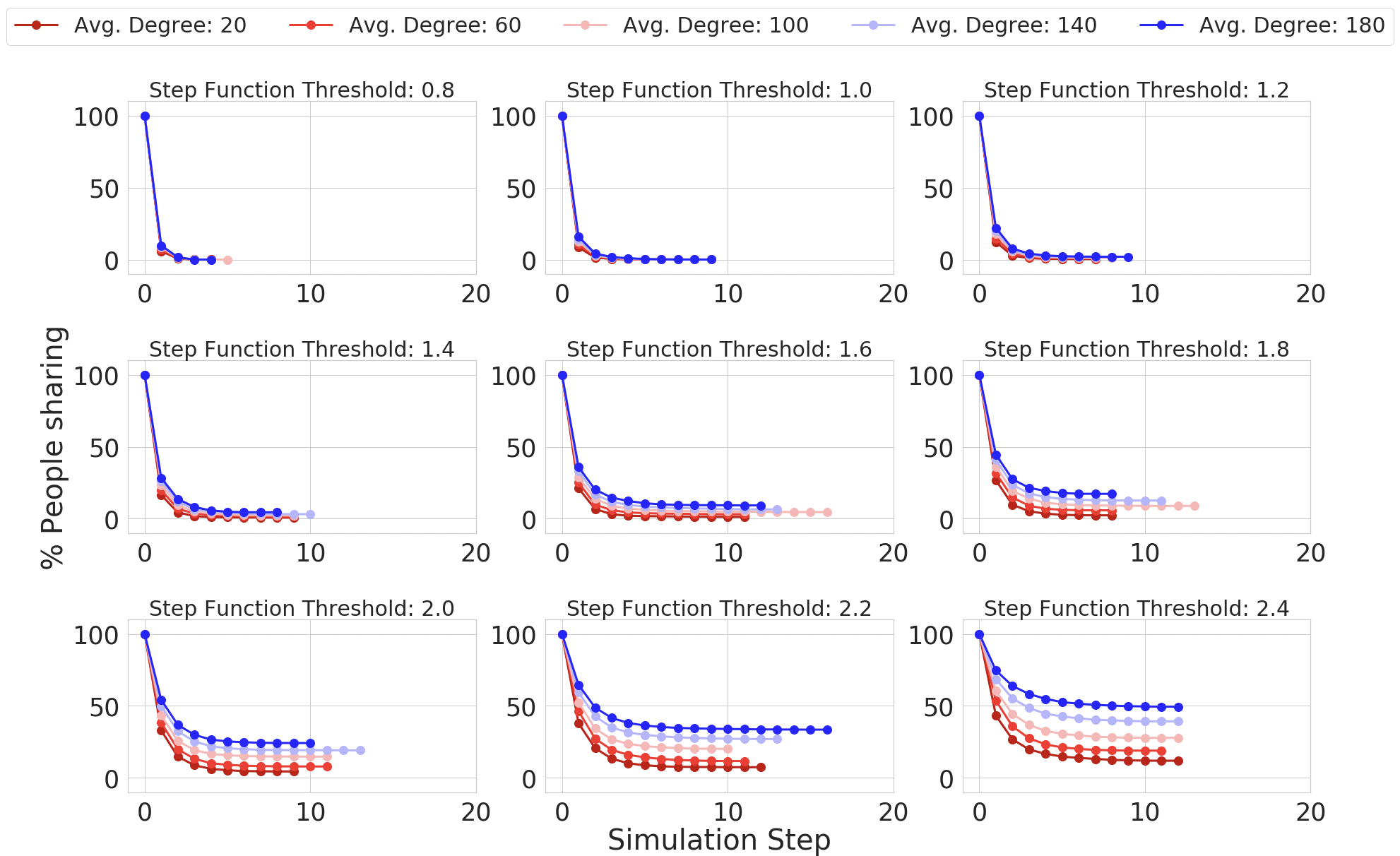}
  \caption{Negative Unit Step Function Simulations run over 9 Barabási-Albert Networks of different average degrees}
  \label{negative_step_barabasi}
\end{figure}

\begin{figure*}[h!]
  \centering
  \includegraphics[width = 0.95\linewidth]{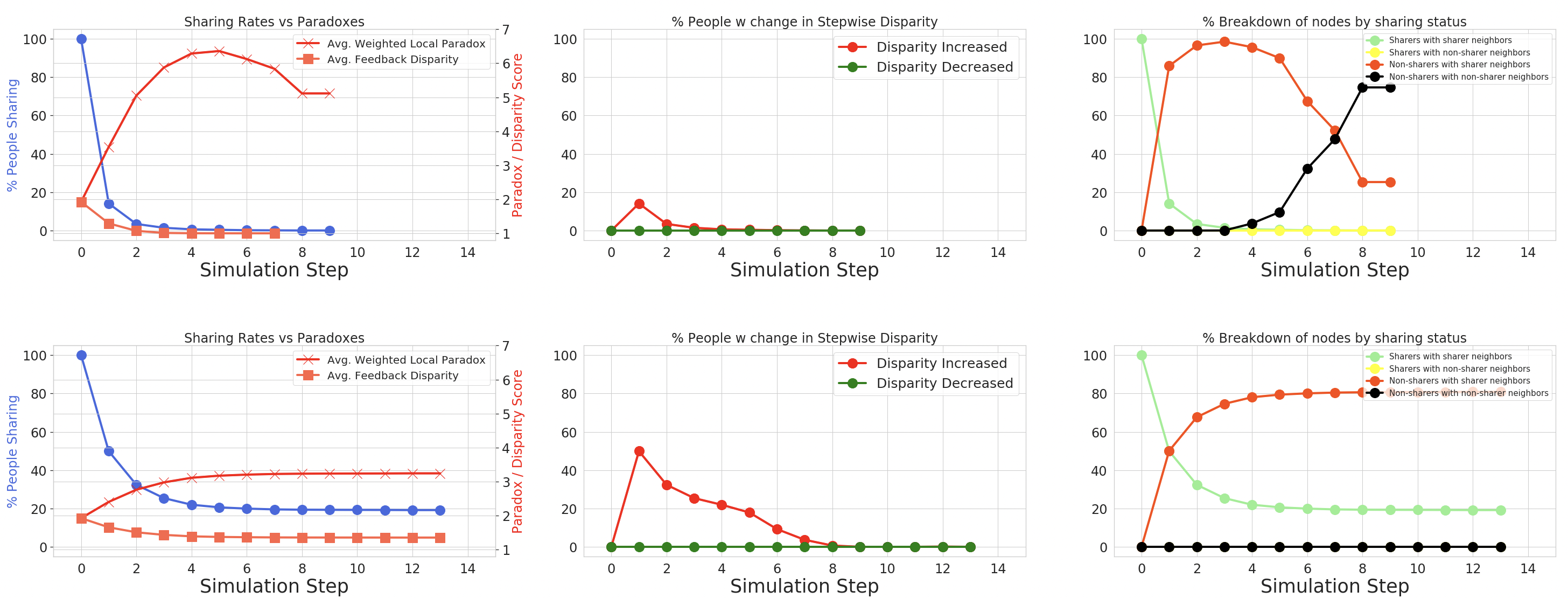}
  \caption{Sharing and Disparity Patterns of a Barabási-Albert graph for negative step function simulation thresholds of 1 (Top) and 2 (Bottom)}
  \label{negative_step_details}
\end{figure*}

As with Erdős–Rényi networks, simulations with step-function thresholds on Barabási-Albert networks show monotonic declines in sharing rates, as shown in figure \ref{negative_step_barabasi}.
However, the decline is more pronounced and generalizes across a wider band of thresholds, with a threshold of 2.5 showing a decline rate nearing 40\% within 15 simulation steps.
As shown in figure \ref{paradox_distributions}, this wideness of effect is explained by Barabási-Albert networks having both a larger average and a larger standard deviation in their local paradox distributions.

To further understand why we observe a gradual asymptotic decline trend, we generate the panel of plots in Figure \ref{negative_step_details}.

\textbf{A point of distinction for these plots:}

\begin{itemize}
\item The average of the weighted local paradox \begin{math}wlp_u(t)=lp_u sb_u(t)\end{math} is computed for all people \begin{math}u\end{math} who have friends who share content, to capture the extent to which sharing bias has magnified the local paradox. 
\item While feedback disparity is equivalent formulaically to the weighted local paradox, however, feedback disparity is only computed for a person \begin{math}u\end{math} who shares content at time \begin{math}t\end{math}, and who also has friends who share content. So its average only includes weighted local paradox scores of a subset of the people included in the average weighted local paradox score.
\end{itemize}

We observe the following in Figure \ref{negative_step_details}:

\begin{enumerate}
\item As sharing rates adjust between steps, we see a percentage of people for whom feedback disparity increases due to the sharing-rate adjustments’ shifting of content creation towards higher-degree connections. These are the people who will drive a decline in the following step. As such, the rate of decline of sharing is congruent with the percentage of people for whom disparity increased in the preceding step.
\item Network-level average feedback disparity declines in tandem with sharing rates. This does not equate to it declining for the group of people that are still sharing, but rather the decline is due to survivor-ship bias of remaining sharers having lower disparities than ones who've churned. Thereby while their disparities go up as shown in the middle plots, the average disparity in the network conversely goes down.
\item Weighted local paradox grows multiples beyond the initial local paradox value in step 0, showing how the initial local paradox in combination with the disparity response function drives an increase in sharing bias towards high degree nodes.
\end{enumerate}

\textbf{Does a low local paradox predict longer surviving nodes?}
If it were a predictor, we would expect a correlation of -1.0 between the local paradox value of a given node, and its terminal sharing step. We compute the Pearson correlation coefficient between the local paradox and terminal sharing step for all nodes in the 10 Barabási-Albert graphs for the negative step threshold 2.0 simulations and find a correlation of -0.69, suggesting that how low a node's local paradox is, is not equivalent to how far a node's sharing survives.

\subsection{Continuous Monotonically Decreasing DRFs}

We have observed that all negative step function simulations showed monotonically decreasing sharing rates that trend towards a non-zero or zero asymptote. We now explore whether, as a class of functions, monotonically decreasing DRFs generally lead to monotonically decreasing sharing rates over time. We do this by looking at other types of monotonically decreasing continuous disparity response functions, such as multiplicative inverse function and negative sloped linear functions (illustrated in Figure \ref{drfs}).

It is trivial to see that, in the first step in simulating a monotonically decreasing DRF, there can be an increase in sharing rates in cases where people with favorable disparities have sufficiently high \begin{math}drf\end{math} values. For example, if someone with local paradox less than 1 shares a million posts instead of 1, then in the first step, it is feasible for overall sharing to go up. Therefore, when making statements about monotonic declines, we are focused on the long-term trend, beyond the first step adjustment.

\begin{figure}[h!]
  \centering
  \includegraphics[width=\linewidth]{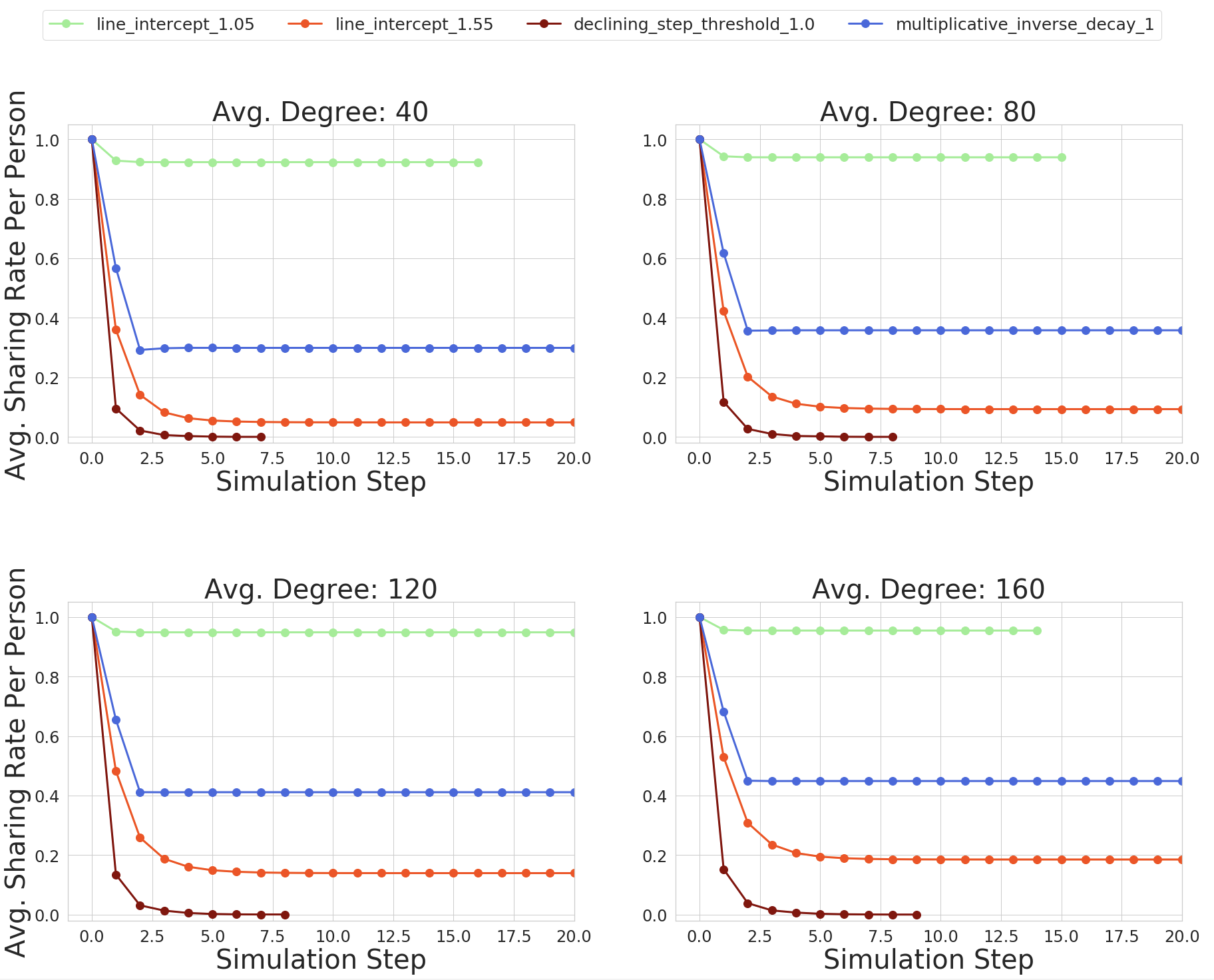}
  \caption{Simulations results of continuous DRFs over 4 Barabási-Albert Networks, for  \begin{math}drf(x) = 1/x\end{math} ({\color{blue} blue}), \begin{math}drf(x) = 1.05 - 0.55x\end{math} ({\color{green} green}) and \begin{math}drf(x) = 1.55 - 0.55x\end{math} ({\color{orange} orange})}
  \label{continuous_barbasi}, 
\end{figure}

\begin{figure*}[h!]
  \centering
  \includegraphics[width = 0.95\linewidth]{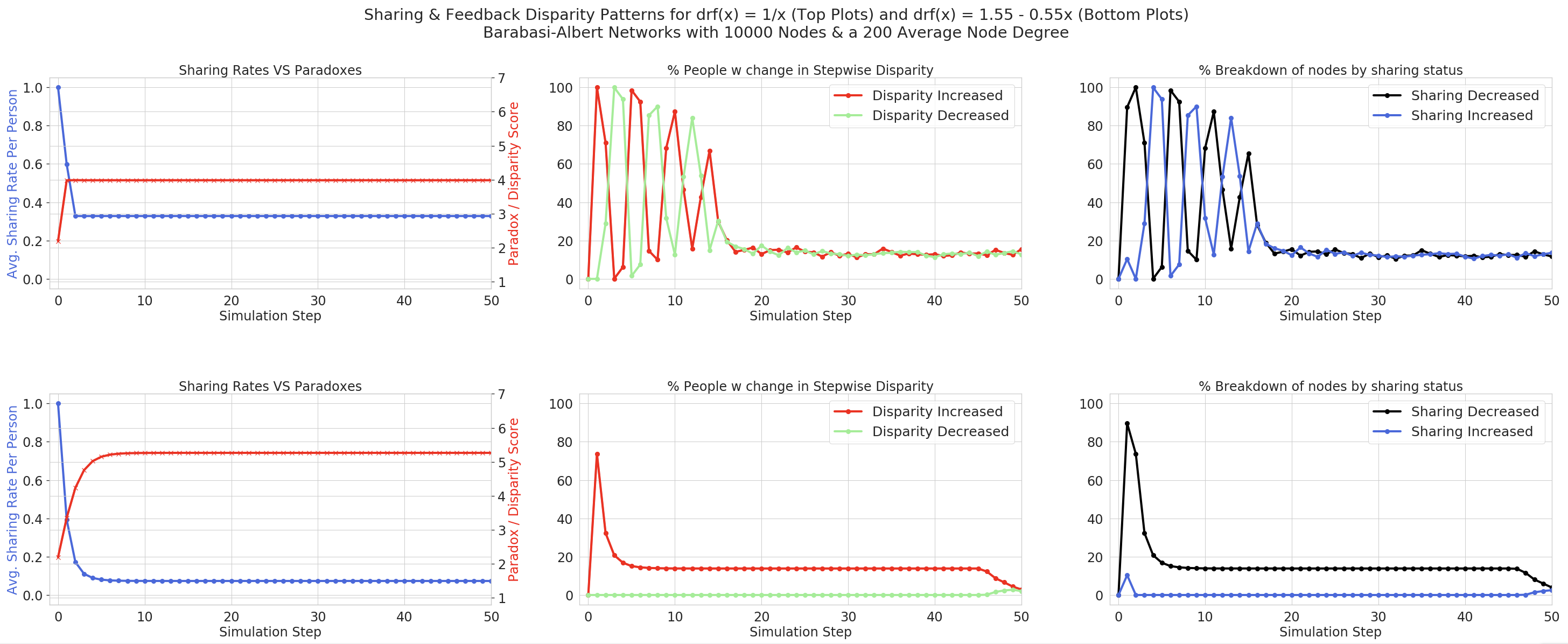}
  \caption{Sharing and Feedback Disparity Patterns of a Barabási-Albert network for \begin{math}drf(x) = 1/x\end{math} (Top) and \begin{math}drf(x) = 1.55 - 0.55x\end{math} (Bottom)}
  \label{1_over_x_details}
\end{figure*}

\textbf{A continuous DRF simulation assumption:} Fractional sharing rates are not discretized in the simulation. This is a more accurate reflection of continuous functions, to capture the scale free effects of sharing trends that would be disrupted by probabilistic churn (i.e. if we were to set initial sharing rate to 100 then a 90\% decline leads to an average of 10 if sampling is applied, rather than a sharing rate of 0 with 90\% chance)

We observe the following key results in figure \ref{continuous_barbasi}:

\begin{itemize}
\item Monotonically decreased DRFs do not uniformly lead to a monotonically decreasing sharing rate. While continuous monotonically decreasing DRFs do overall end-up in a declining state, declines are actually not monotonic. This is evident in the network with a 20 degree average in figure \ref{continuous_barbasi}, where the multiplicative inverse DRF drives a sharing rate increase between steps 2 and 3. This can happen through the occasional churn of a friend \begin{math}v\end{math} with higher feedback than friend \begin{math}u\end{math}, due to \begin{math}v\end{math}'s friends having even higher feedback, without \begin{math}u\end{math} churning, leading to a decrease in feedback disparities, which can bring sharing rates up.
\item Continuous monotonically decreasing DRFs can reach non-zero equilibrium states, rather than declining towards zero.

\end{itemize}

To better understand why continuous DRFs converge around non-zero equilibrium points, we plot sharing rates, average weighted local paradox, average disparity and how these values move up and down for people over a simulation run for a Barabási Albert Network with 1000 nodes and a 200 average degree per node, in figure \ref{1_over_x_details}.

As shown in figure \ref{1_over_x_details}, we observe that convergence does not necessarily happen via a gradually slowing decline in sharing rate to an asymptote but also through a narrowing oscillatory convergence around an equilibrium point, as is the case for \begin{math}drf(x) = \frac{1}{x}\end{math}. 

\subsection{Convex DRFs}

\begin{figure}[h]
  \centering
  \includegraphics[width=\linewidth]{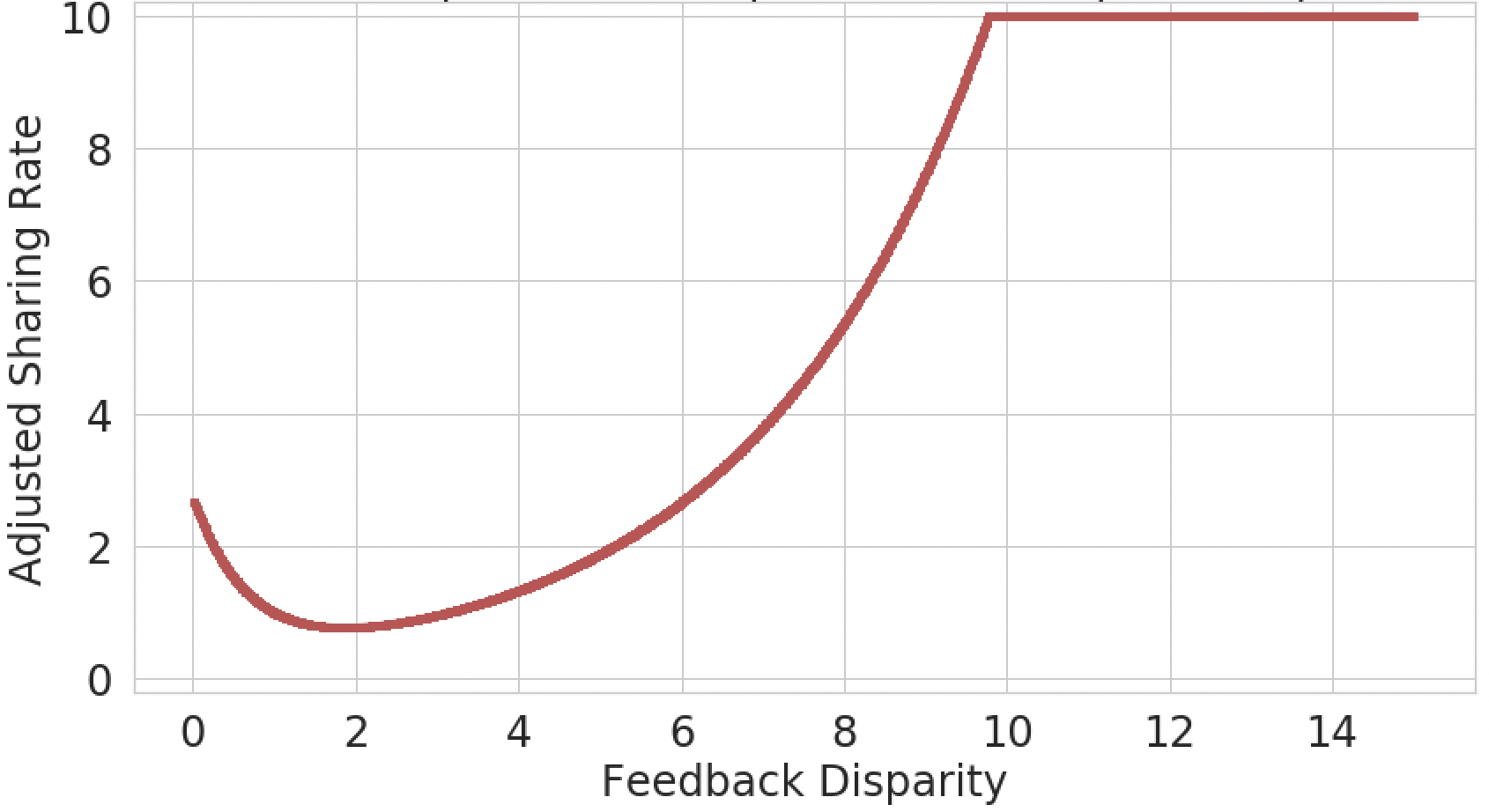}
  \caption{Convex DRF \begin{math}drf(x) = min(\frac{e^{1-1.5X} + e^{0.35X - 1}}{e^{1} + e^{-1}}, 10)\end{math}}
  \label{drf_convex}
\end{figure}

In this section, we look at overall sharing rates over time in the case of a convex function expressed by the convex DRF in equation \ref{convex_drf_equation}, whose curve is shown in figure \ref{drf_convex}.

\begin{figure}[h]
  \centering
  \includegraphics[width=\linewidth]{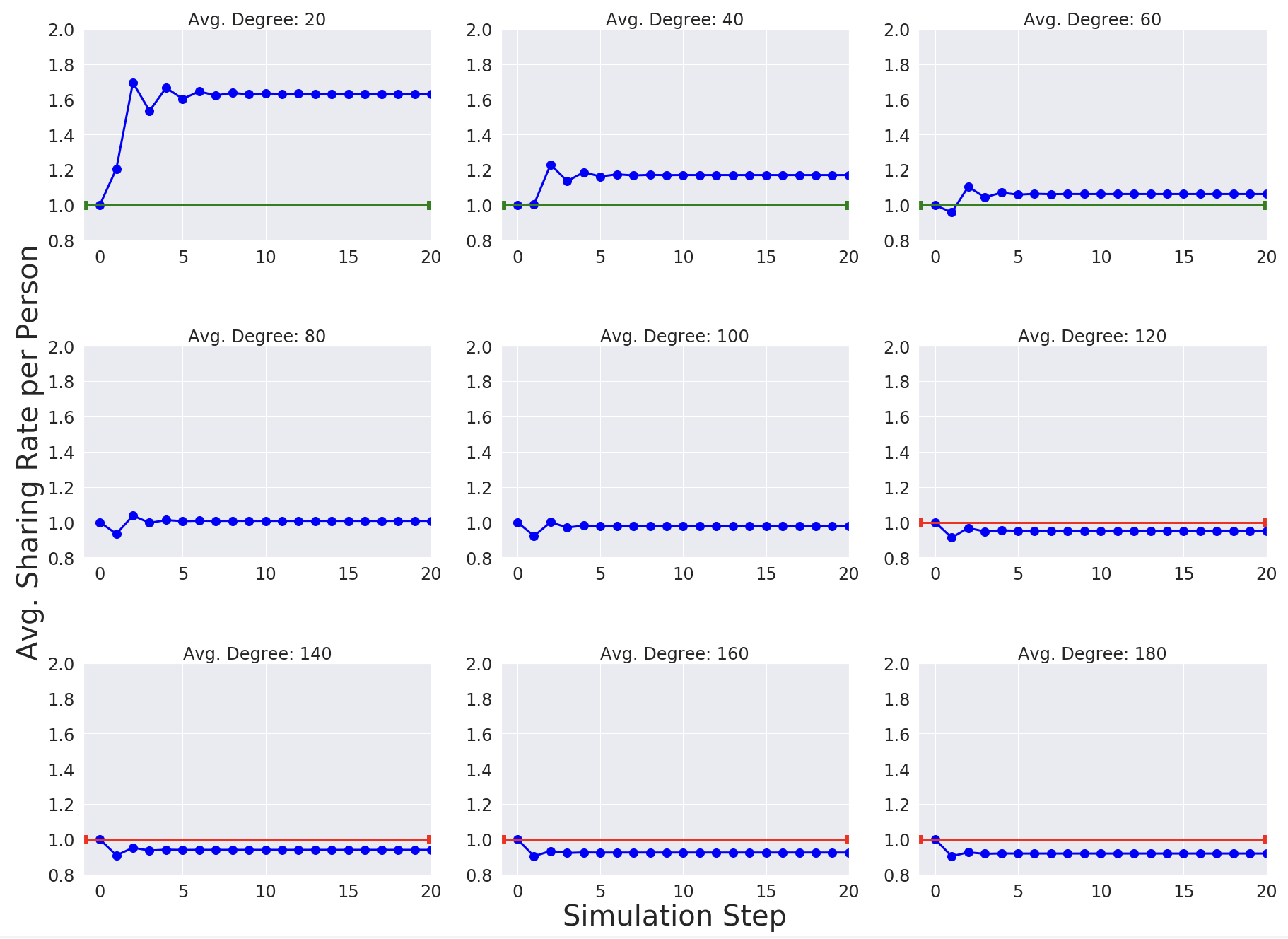}
  \caption{Simulation results over 9 Barabási-Albert  Networks of different average degrees of the convex function \begin{math}drf(x) = min(\frac{e^{1-1.5X} + e^{0.35X - 1}}{e^{1} + e^{-1}}, 10)\end{math}}
  \label{convex_barbasi}
\end{figure}

We observe in figure \ref{convex_barbasi} that:

\begin{itemize}
    \item Convex DRFs can lead to sustained or increased sharing rates.
    \item Direction of impact of convex DRFs is network dependent. The same Convex DRF can trigger an increase in sharing rates for one network and a decrease for another. \textbf{This is a particularly stark key conclusion that can prove value for real world use cases.}
    \item Whether it causes an increase or decrease, there is always an equilibrium point for any of the networks simulated, rather than sharing trending indefinitely up or down to 0.
\end{itemize}

\subsection{Positive Unit Step Function DRFs}
\begin{figure}[h]
  \centering
  \includegraphics[width=\linewidth]{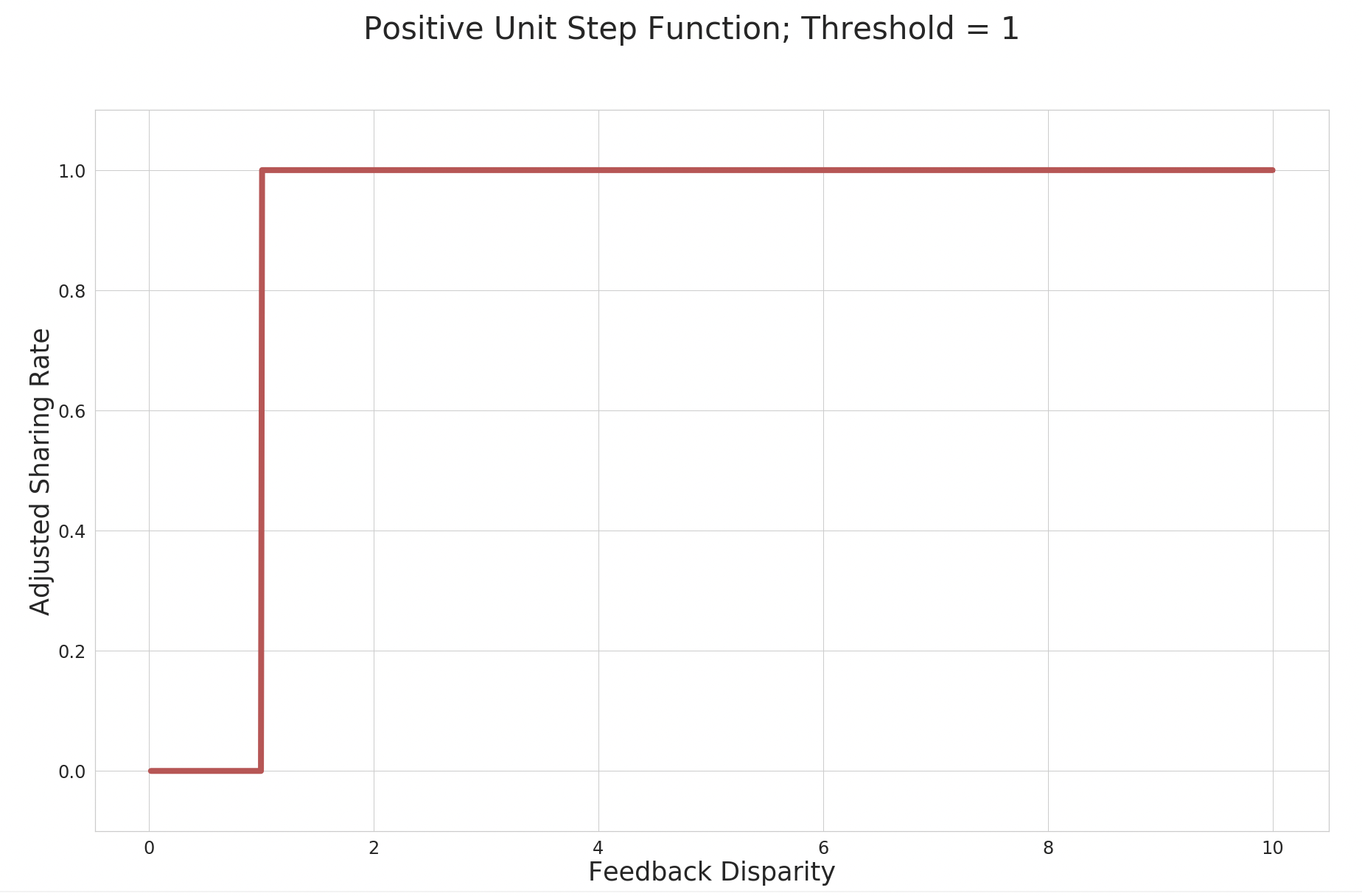}
  \caption{Positive Unit Step Function DRF}
  \label{drf_types_positive_step}
\end{figure}

So far we have shown that monotonically decreasing disparity response functions lead to reductions in overall sharing rates, whereas convex functions can lead to a growth in sharing rates over time depending on the network it applies to. 
In this section, we alternatively simulate a positive  unit step function where people only share if their feedback disparity exceeds a threshold, i.e. people are actually discouraged to share the more popular they are than their friends, in terms of the feedback volume that they receive. The simulated DRF is shown in figure \ref{drf_types_positive_step}.

\begin{figure*}[h]
  \centering
  \includegraphics[width = 0.95\linewidth]{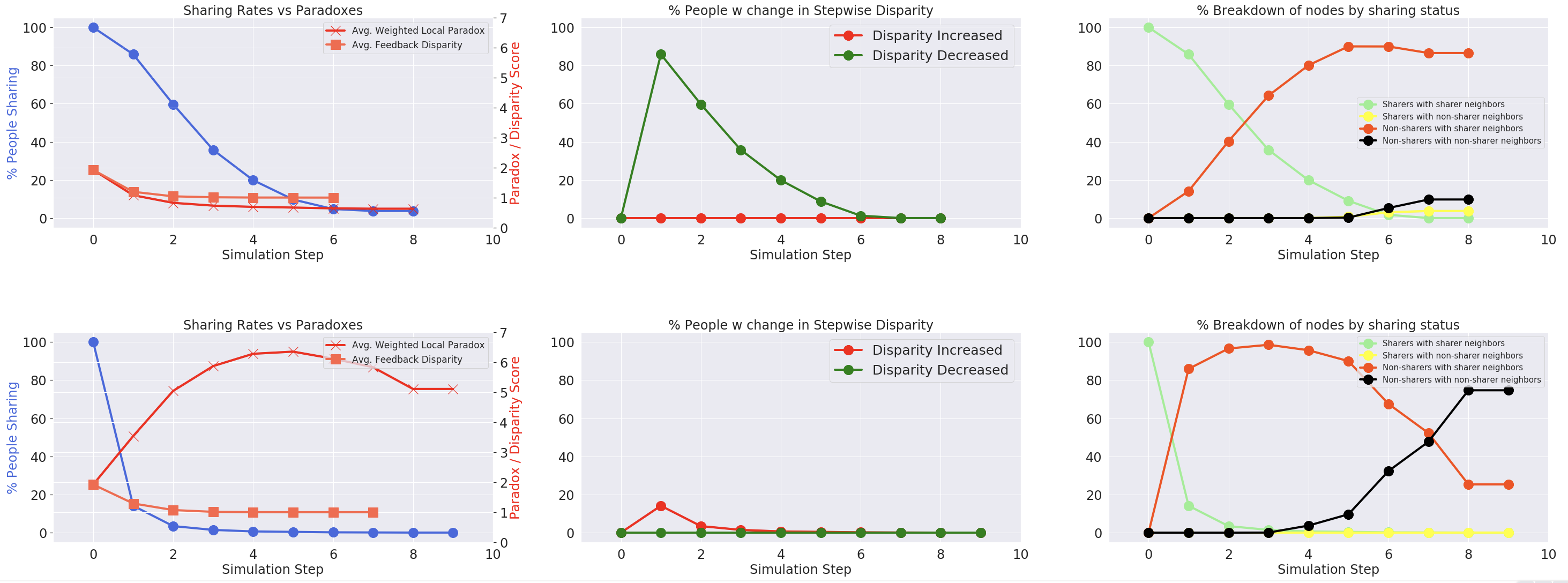}
  \caption{Sharing and Feedback Disparity Patterns of a Barabási-Albert network for a threshold 1 positive step function (Top) and a threshold 1 negative step function (Bottom)}
  \label{positive_step_details}
\end{figure*}

We observe a somewhat counter intuitive simulation result, with the positive step function also leading to a gradual decline in sharing rates over time. This decline is because when people with a lot of feedback stop sharing, feedback disparities start to decrease in the network, and some people who were encouraged to share are now discouraged from doing so. We show this effect in figure \ref{positive_step_details}. Note how, opposite to negative step functions, the positive step function threshold leads to nodes experiencing a decrease in their disparity between steps.

\subsection{Interaction of a Disparity Threshold with an Activity Threshold}

\begin{figure}[h!]
  \centering
  \includegraphics[width=\linewidth]{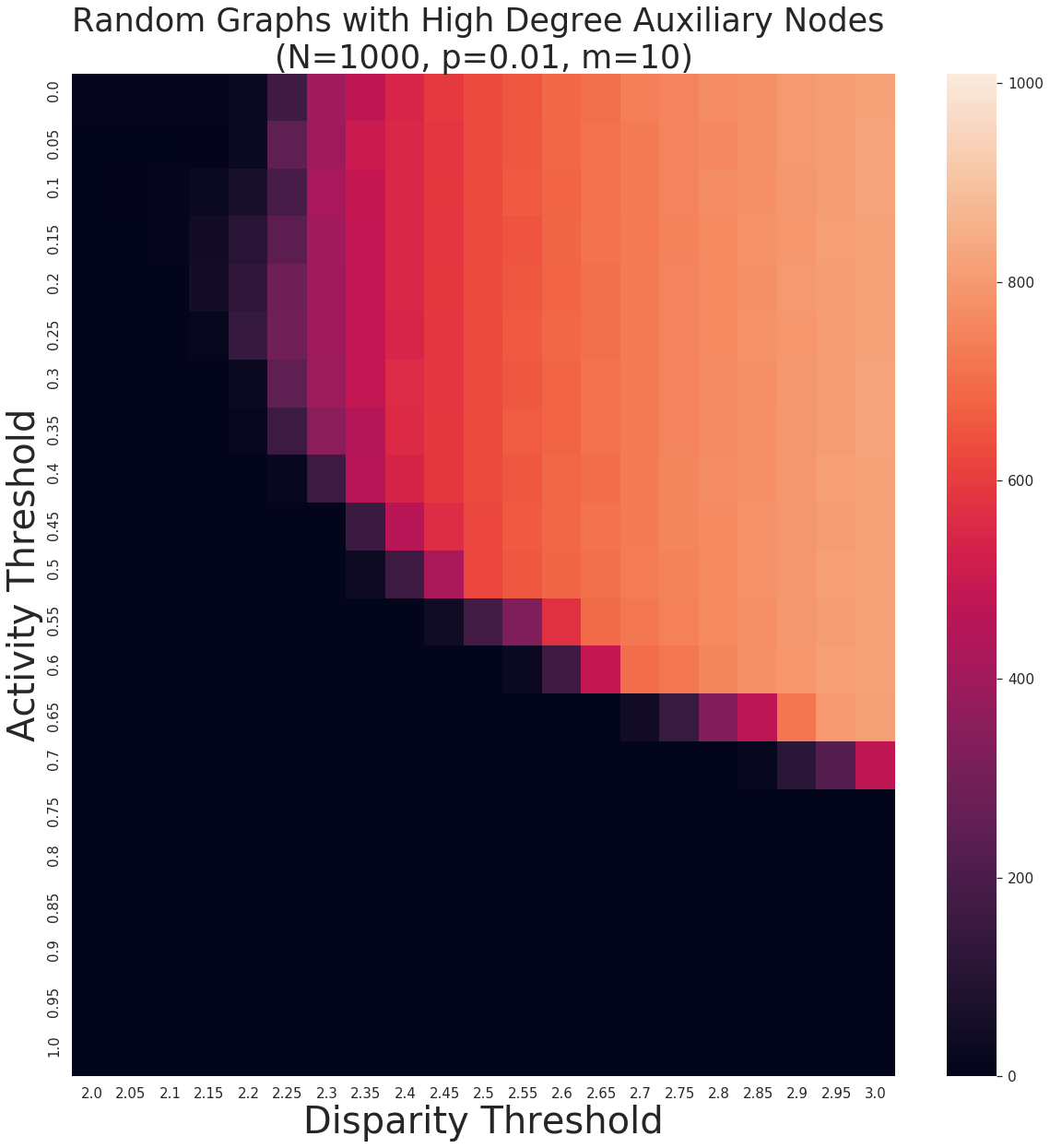}
  \caption{Terminal sharing rates for Erdös-Rényi simulations with different activity and disparity thresholds}
  \label{heatmap_activity_thresholds_erdos}
\end{figure}

As summarized in section 2, a relevant extension is combining disparity response functions with threshold functions of collective behavior.
In figure \ref{heatmap_activity_thresholds_erdos}, we consider the case of Erdös-Rényi random graphs with 1000 nodes and link probability 0.01. 
To such a random graph, we add 10 additional nodes, and each node in the original random graph connects at random to one of these 10 additional nodes. 
We average over 50 realizations of this random graph-generation process for each combination of disparity and activity thresholds. 
We find, for example when the disparity threshold is 2.2, that no one shares in the long term when the activity threshold is 0, but that some sharing persists at intermediate values of the activity threshold. 
This is because the high degree ``auxiliary” nodes tend to churn from sharing in early stages of the simulation (due to the activity threshold), reducing values of the disparity for the remaining nodes.

\section{Concluding Remarks}
\subsection{Summary}
In this paper, we proposed and analyzed a model for studying the role that the friendship paradox can play in a social network where people can contribute content, their friends can give it feedback, and everyone sees the feedback given by these friends.
By not factoring in engagement bias in our initial model, our model focused entirely on how two variables can drive long-term sharing trajectories:
\begin{enumerate}
    \item network structure, which results in local structural friendship paradoxes 
    \item how people respond to social comparison, encoded in a disparity response function (DRF)
\end{enumerate}
We studied Erdős–Rényi and Barabási-Albert networks, and different types of DRFs, including negative and positive step functions, continuous monotonically decreasing DRFs and continuous convex DRFs.

Our key findings are:
\begin{itemize}
\item All DRFs lead to a sharing equilibrium point where sharing rates do not increase or decrease further.
\item  Monotonicity of DRFs was overall predictive of a gradual decline in sharing rates, for both monotonically increasing and decreasing DRFs.
\item Convex DRFs can sustain or grow sharing rates.
\item The direction of impact of convex DRFs is network dependent. The same convex disparity response function can lead to a growth in sharing rates in one network and a decline in another.
\end{itemize}
These findings suggest that the opportunity for sustainable sharing trajectories lies in complementary types of sharing incentives, with some incentives adapting to unfavorable social comparisons and others benefiting from favorable ones, and with some degree of tailoring of such incentives to the network this applies to.

\subsection{Practical implications for real-world networks}

To put forward an initial foundation for a more complete network structure informed social sharing theory, we have made multiple simplifications in this research. 
We have chosen to focus on synthetic networks, and we did not attempt to alter engagement rate biases. 
Also, we have not factored in the multitude of factors that can modulate social sharing propensities in the real world, including effort involved in sharing, activity thresholds, viewer preferences, and other factors that can play a role in social sharing \cite{oliveira2020people, shu2011people, lee2013people, berger2012facebook}. 
With that in mind, more work is needed to understand what our model can say about real-world online networks.

Despite that,due to the universality of the friendship paradox as a structural bias of social networks that can have an incremental impact on engagement biases and other sharing effects, our methods can be are useful as a tool for sustaining sharing in online social networks, even in the presence of adverse social comparison effects.

\textbf{There are 3 key opportunity areas for online networks}
\begin{enumerate}
    \item \textbf{Engagement:} Networks can run causal analyses to infer their disparity response functions, and use that to up-level exposure for people with less favorable responses to their perceived feedback disparities.
    \item \textbf{Network Structure:} given that different networks respond differently to the same disparity response function, networks can target growing connections that work best with their respective disparity response functions.
    \item \textbf{Product Design:} The choice of feedback on content being widely visible creates a trade-off between providing a useful signal for content creators and consumers, and triggering adverse social comparison effects impacting sharing rates. Adopting systems where feedback visibility is removed or constrained to balance these trade-offs can provide value. Prior research has made this point in relation to social comparisons on Facebook \cite{burke2020social}.
\end{enumerate}

Further, our model can be useful for studying other  types of friendship paradox induced social comparisons in real world offline networks, in cases where such comparisons are known to mediate behavioral responses.

\subsection{Next steps and research opportunities}
There are multiple opportunities for future research. Of note are;
\begin{enumerate}
\item Understanding how the model behaves in real online networks?
\item Given the convex DRFs result, what happens to a network when we fix degree distributions and alter assortativity? or fix both and alter transsortativity? \cite{ngo2020transsortative}.
\item How does the model behave when we incorporate other sharing drivers and incentives? 
\end{enumerate}

\bibliographystyle{ACM-Reference-Format}
\bibliography{sample-base}


\begin{thebibliography}{24}

%%% the Web address.

\ifx \showCODEN    \undefined \def \showCODEN     #1{\unskip}     \fi
\ifx \showDOI      \undefined \def \showDOI       #1{#1}\fi
\ifx \showISBNx    \undefined \def \showISBNx     #1{\unskip}     \fi
\ifx \showISBNxiii \undefined \def \showISBNxiii  #1{\unskip}     \fi
\ifx \showISSN     \undefined \def \showISSN      #1{\unskip}     \fi
\ifx \showLCCN     \undefined \def \showLCCN      #1{\unskip}     \fi
\ifx \shownote     \undefined \def \shownote      #1{#1}          \fi
\ifx \showarticletitle \undefined \def \showarticletitle #1{#1}   \fi
\ifx \showURL      \undefined \def \showURL       {\relax}        \fi
% invisible to TeX
\providecommand\bibfield[2]{#2}
\providecommand\bibinfo[2]{#2}
\providecommand\natexlab[1]{#1}
\providecommand\showeprint[2][]{arXiv:#2}

\bibitem[Barab{\'a}si and Albert(1999)]%
        {barabasi1999emergence}
\bibfield{author}{\bibinfo{person}{Albert-L{\'a}szl{\'o} Barab{\'a}si} {and}
  \bibinfo{person}{R{\'e}ka Albert}.} \bibinfo{year}{1999}\natexlab{}.
\newblock \showarticletitle{Emergence of scaling in random networks}.
\newblock \bibinfo{journal}{\emph{science}} \bibinfo{volume}{286},
  \bibinfo{number}{5439} (\bibinfo{year}{1999}), \bibinfo{pages}{509--512}.
\newblock


\bibitem[Berger and Buechel(2012)]%
        {berger2012facebook}
\bibfield{author}{\bibinfo{person}{Jonah~A Berger} {and} \bibinfo{person}{Eva
  Buechel}.} \bibinfo{year}{2012}\natexlab{}.
\newblock \showarticletitle{Facebook therapy? Why do people share self-relevant
  content online?}
\newblock \bibinfo{journal}{\emph{Why Do People Share Self-Relevant Content
  Online}} (\bibinfo{year}{2012}).
\newblock


\bibitem[Bollen et~al\mbox{.}(2017)]%
        {bollen2017happiness}
\bibfield{author}{\bibinfo{person}{Johan Bollen}, \bibinfo{person}{Bruno
  Gon{\c{c}}alves}, \bibinfo{person}{Ingrid van~de Leemput}, {and}
  \bibinfo{person}{Guangchen Ruan}.} \bibinfo{year}{2017}\natexlab{}.
\newblock \showarticletitle{The happiness paradox: your friends are happier
  than you}.
\newblock \bibinfo{journal}{\emph{EPJ Data Science}} \bibinfo{volume}{6},
  \bibinfo{number}{1} (\bibinfo{year}{2017}), \bibinfo{pages}{4}.
\newblock


\bibitem[Breiger and Pattison(1986)]%
        {breiger1986cumulated}
\bibfield{author}{\bibinfo{person}{Ronald~L Breiger} {and}
  \bibinfo{person}{Philippa~E Pattison}.} \bibinfo{year}{1986}\natexlab{}.
\newblock \showarticletitle{Cumulated social roles: The duality of persons and
  their algebras}.
\newblock \bibinfo{journal}{\emph{Social networks}} \bibinfo{volume}{8},
  \bibinfo{number}{3} (\bibinfo{year}{1986}), \bibinfo{pages}{215--256}.
\newblock


\bibitem[Burke et~al\mbox{.}(2020)]%
        {burke2020social}
\bibfield{author}{\bibinfo{person}{Moira Burke}, \bibinfo{person}{Justin
  Cheng}, {and} \bibinfo{person}{Bethany de Gant}.}
  \bibinfo{year}{2020}\natexlab{}.
\newblock \showarticletitle{Social comparison and Facebook: Feedback,
  positivity, and opportunities for comparison}. In
  \bibinfo{booktitle}{\emph{Proceedings of the 2020 CHI conference on human
  factors in computing systems}}. \bibinfo{pages}{1--13}.
\newblock


\bibitem[Eom and Jo(2014)]%
        {eom2014generalized}
\bibfield{author}{\bibinfo{person}{Young-Ho Eom} {and}
  \bibinfo{person}{Hang-Hyun Jo}.} \bibinfo{year}{2014}\natexlab{}.
\newblock \showarticletitle{Generalized friendship paradox in complex networks:
  The case of scientific collaboration}.
\newblock \bibinfo{journal}{\emph{Scientific reports}} \bibinfo{volume}{4},
  \bibinfo{number}{1} (\bibinfo{year}{2014}), \bibinfo{pages}{1--6}.
\newblock


\bibitem[Erd{\H{o}}s et~al\mbox{.}(1960)]%
        {erdHos1960evolution}
\bibfield{author}{\bibinfo{person}{Paul Erd{\H{o}}s},
  \bibinfo{person}{Alfr{\'e}d R{\'e}nyi}, {et~al\mbox{.}}}
  \bibinfo{year}{1960}\natexlab{}.
\newblock \showarticletitle{On the evolution of random graphs}.
\newblock \bibinfo{journal}{\emph{Publ. Math. Inst. Hung. Acad. Sci}}
  \bibinfo{volume}{5}, \bibinfo{number}{1} (\bibinfo{year}{1960}),
  \bibinfo{pages}{17--60}.
\newblock


\bibitem[Feld(1991)]%
        {feld1991your}
\bibfield{author}{\bibinfo{person}{Scott~L Feld}.}
  \bibinfo{year}{1991}\natexlab{}.
\newblock \showarticletitle{Why your friends have more friends than you do}.
\newblock \bibinfo{journal}{\emph{American journal of sociology}}
  \bibinfo{volume}{96}, \bibinfo{number}{6} (\bibinfo{year}{1991}),
  \bibinfo{pages}{1464--1477}.
\newblock


\bibitem[Geng et~al\mbox{.}(2019)]%
        {geng2019sentinel}
\bibfield{author}{\bibinfo{person}{Jiachen Geng}, \bibinfo{person}{Yuanxi Li},
  \bibinfo{person}{Zili Zhang}, {and} \bibinfo{person}{Li Tao}.}
  \bibinfo{year}{2019}\natexlab{}.
\newblock \showarticletitle{Sentinel nodes identification for infectious
  disease surveillance on temporal social networks}. In
  \bibinfo{booktitle}{\emph{IEEE/WIC/ACM International Conference on Web
  Intelligence}}. \bibinfo{pages}{493--499}.
\newblock


\bibitem[Higham(2019)]%
        {higham2019centrality}
\bibfield{author}{\bibinfo{person}{Desmond~J Higham}.}
  \bibinfo{year}{2019}\natexlab{}.
\newblock \showarticletitle{Centrality-friendship paradoxes: when our friends
  are more important than us}.
\newblock \bibinfo{journal}{\emph{Journal of Complex Networks}}
  (\bibinfo{year}{2019}).
\newblock


\bibitem[Hodas et~al\mbox{.}(2013)]%
        {hodas2013friendship}
\bibfield{author}{\bibinfo{person}{Nathan Hodas}, \bibinfo{person}{Farshad
  Kooti}, {and} \bibinfo{person}{Kristina Lerman}.}
  \bibinfo{year}{2013}\natexlab{}.
\newblock \showarticletitle{Friendship paradox redux: Your friends are more
  interesting than you}. In \bibinfo{booktitle}{\emph{Proceedings of the
  International AAAI Conference on Web and Social Media}},
  Vol.~\bibinfo{volume}{7}. \bibinfo{pages}{225--233}.
\newblock


\bibitem[Jackson(2019)]%
        {jackson2019friendship}
\bibfield{author}{\bibinfo{person}{Matthew~O Jackson}.}
  \bibinfo{year}{2019}\natexlab{}.
\newblock \showarticletitle{The friendship paradox and systematic biases in
  perceptions and social norms}.
\newblock \bibinfo{journal}{\emph{Journal of Political Economy}}
  \bibinfo{volume}{127}, \bibinfo{number}{2} (\bibinfo{year}{2019}),
  \bibinfo{pages}{777--818}.
\newblock


\bibitem[Kooti et~al\mbox{.}(2014)]%
        {kooti2014network}
\bibfield{author}{\bibinfo{person}{Farshad Kooti}, \bibinfo{person}{Nathan~O
  Hodas}, {and} \bibinfo{person}{Kristina Lerman}.}
  \bibinfo{year}{2014}\natexlab{}.
\newblock \showarticletitle{Network weirdness: Exploring the origins of network
  paradoxes}. In \bibinfo{booktitle}{\emph{Eighth International AAAI Conference
  on Weblogs and Social Media}}.
\newblock


\bibitem[Lee et~al\mbox{.}(2019)]%
        {lee2019impact}
\bibfield{author}{\bibinfo{person}{Eun Lee}, \bibinfo{person}{Sungmin Lee},
  \bibinfo{person}{Young-Ho Eom}, \bibinfo{person}{Petter Holme}, {and}
  \bibinfo{person}{Hang-Hyun Jo}.} \bibinfo{year}{2019}\natexlab{}.
\newblock \showarticletitle{Impact of perception models on friendship paradox
  and opinion formation}.
\newblock \bibinfo{journal}{\emph{Physical Review E}} \bibinfo{volume}{99},
  \bibinfo{number}{5} (\bibinfo{year}{2019}), \bibinfo{pages}{052302}.
\newblock


\bibitem[Lee et~al\mbox{.}(2013)]%
        {lee2013people}
\bibfield{author}{\bibinfo{person}{Haein Lee}, \bibinfo{person}{Hyejin Park},
  {and} \bibinfo{person}{Jinwoo Kim}.} \bibinfo{year}{2013}\natexlab{}.
\newblock \showarticletitle{Why do people share their context information on
  Social Network Services? A qualitative study and an experimental study on
  users' behavior of balancing perceived benefit and risk}.
\newblock \bibinfo{journal}{\emph{International Journal of Human-Computer
  Studies}} \bibinfo{volume}{71}, \bibinfo{number}{9} (\bibinfo{year}{2013}),
  \bibinfo{pages}{862--877}.
\newblock


\bibitem[Macy and Evtushenko(2020)]%
        {macy2020threshold}
\bibfield{author}{\bibinfo{person}{Michael~W Macy} {and} \bibinfo{person}{Anna
  Evtushenko}.} \bibinfo{year}{2020}\natexlab{}.
\newblock \showarticletitle{Threshold models of collective behavior ii: The
  predictability paradox and spontaneous instigation}.
\newblock \bibinfo{journal}{\emph{Sociological Science}}  \bibinfo{volume}{7}
  (\bibinfo{year}{2020}), \bibinfo{pages}{628--648}.
\newblock


\bibitem[Nettasinghe and Krishnamurthy(2019)]%
        {nettasinghe2019your}
\bibfield{author}{\bibinfo{person}{Buddhika Nettasinghe} {and}
  \bibinfo{person}{Vikram Krishnamurthy}.} \bibinfo{year}{2019}\natexlab{}.
\newblock \showarticletitle{“What do your friends think?”: Efficient
  polling methods for networks using friendship paradox}.
\newblock \bibinfo{journal}{\emph{IEEE Transactions on Knowledge and Data
  Engineering}} \bibinfo{volume}{33}, \bibinfo{number}{3}
  (\bibinfo{year}{2019}), \bibinfo{pages}{1291--1305}.
\newblock


\bibitem[Nettasinghe et~al\mbox{.}(2019)]%
        {nettasinghe2019diffusion}
\bibfield{author}{\bibinfo{person}{Buddhika Nettasinghe},
  \bibinfo{person}{Vikram Krishnamurthy}, {and} \bibinfo{person}{Kristina
  Lerman}.} \bibinfo{year}{2019}\natexlab{}.
\newblock \showarticletitle{Diffusion in social networks: Effects of monophilic
  contagion, friendship paradox, and reactive networks}.
\newblock \bibinfo{journal}{\emph{IEEE Transactions on Network Science and
  Engineering}} \bibinfo{volume}{7}, \bibinfo{number}{3}
  (\bibinfo{year}{2019}), \bibinfo{pages}{1121--1132}.
\newblock


\bibitem[Ngo et~al\mbox{.}(2020)]%
        {ngo2020transsortative}
\bibfield{author}{\bibinfo{person}{Shin-Chieng Ngo}, \bibinfo{person}{Allon~G
  Percus}, \bibinfo{person}{Keith Burghardt}, {and} \bibinfo{person}{Kristina
  Lerman}.} \bibinfo{year}{2020}\natexlab{}.
\newblock \showarticletitle{The transsortative structure of networks}.
\newblock \bibinfo{journal}{\emph{Proceedings of the Royal Society A}}
  \bibinfo{volume}{476}, \bibinfo{number}{2237} (\bibinfo{year}{2020}),
  \bibinfo{pages}{20190772}.
\newblock


\bibitem[Oliveira et~al\mbox{.}(2020)]%
        {oliveira2020people}
\bibfield{author}{\bibinfo{person}{Tiago Oliveira}, \bibinfo{person}{Benedita
  Araujo}, {and} \bibinfo{person}{Carlos Tam}.}
  \bibinfo{year}{2020}\natexlab{}.
\newblock \showarticletitle{Why do people share their travel experiences on
  social media?}
\newblock \bibinfo{journal}{\emph{Tourism Management}}  \bibinfo{volume}{78}
  (\bibinfo{year}{2020}), \bibinfo{pages}{104041}.
\newblock


\bibitem[Pires et~al\mbox{.}(2017)]%
        {pires2017friendship}
\bibfield{author}{\bibinfo{person}{Mathias~M Pires}, \bibinfo{person}{Flavia~MD
  Marquitti}, {and} \bibinfo{person}{Paulo~R Guimaraes~Jr}.}
  \bibinfo{year}{2017}\natexlab{}.
\newblock \showarticletitle{The friendship paradox in species-rich ecological
  networks: Implications for conservation and monitoring}.
\newblock \bibinfo{journal}{\emph{Biological conservation}}
  \bibinfo{volume}{209} (\bibinfo{year}{2017}), \bibinfo{pages}{245--252}.
\newblock


\bibitem[Scissors et~al\mbox{.}(2016)]%
        {scissors2016s}
\bibfield{author}{\bibinfo{person}{Lauren Scissors}, \bibinfo{person}{Moira
  Burke}, {and} \bibinfo{person}{Steven Wengrovitz}.}
  \bibinfo{year}{2016}\natexlab{}.
\newblock \showarticletitle{What's in a Like? Attitudes and behaviors around
  receiving Likes on Facebook}. In \bibinfo{booktitle}{\emph{Proceedings of the
  19th acm conference on computer-supported cooperative work \& social
  computing}}. \bibinfo{pages}{1501--1510}.
\newblock


\bibitem[Shu and Chuang(2011)]%
        {shu2011people}
\bibfield{author}{\bibinfo{person}{Wesley Shu} {and} \bibinfo{person}{Yu-Hao
  Chuang}.} \bibinfo{year}{2011}\natexlab{}.
\newblock \showarticletitle{Why people share knowledge in virtual communities}.
\newblock \bibinfo{journal}{\emph{Social Behavior and Personality: an
  international journal}} \bibinfo{volume}{39}, \bibinfo{number}{5}
  (\bibinfo{year}{2011}), \bibinfo{pages}{671--690}.
\newblock


\bibitem[Yan et~al\mbox{.}(2022)]%
        {yan2022does}
\bibfield{author}{\bibinfo{person}{Shen Yan}, \bibinfo{person}{Kristen~M
  Altenburger}, \bibinfo{person}{Yi-Chia Wang}, {and} \bibinfo{person}{Justin
  Cheng}.} \bibinfo{year}{2022}\natexlab{}.
\newblock \showarticletitle{What Does Perception Bias on Social Networks Tell
  Us About Friend Count Satisfaction?}. In
  \bibinfo{booktitle}{\emph{Proceedings of the ACM Web Conference 2022}}.
  \bibinfo{pages}{2687--2695}.
\newblock


\end{thebibliography}

\appendix

\section{Appendix}

\subsection{Illustrative examples}

\subsubsection{Disparity and Activity Thresholds Applied to FFN}

\begin{figure}[h]
  \centering
  \includegraphics[width=\linewidth]{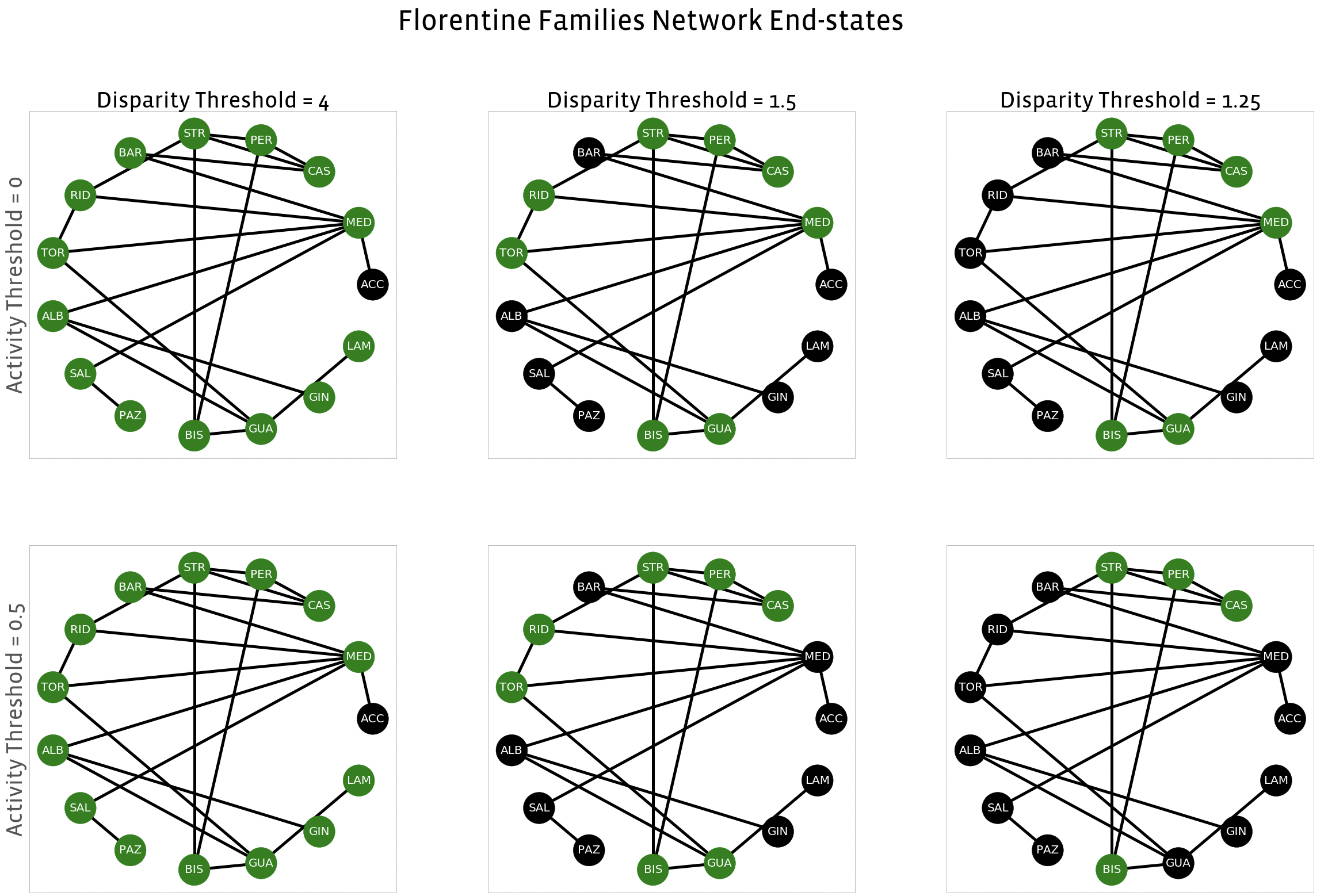}
  \caption{Illustrating how the FIT model leads to different end-state sharing rates for different disparity and activity thresholds}
  \label{florentine_w_activity_threshold}
\end{figure}

In figure \ref{florentine_w_activity_threshold}, we show implications of applying different activity and disparity thresholds on end-state results of simulations of our model. When there is no activity threshold, nodes that share in the long term (shown in green) are those that are most central in the network. For example, the Medici, Guadagni, and Strozzi continue sharing even with an aggressive disparity threshold of 1. As these nodes do not experience a local structural friendship paradox (i.e., their degree is higher than the average degree of their friends), nor do they experience a feedback disparity at any point during the simulation. In contrast, the Castellani do not experience a local paradox, but they do experience a feedback disparity once the Peruzzi and Barbadori stop sharing. Therefore, they do not share in the long term. Once there is an activity threshold, the relationship between centrality and long-term sharing is not as tight. With a disparity threshold of 1.5 and an activity threshold of 0.5 for instance, the Medici eventually stop sharing. This is because, although the Medici themselves have high degree in the network, they tend to link to families that have relatively low degree, such as the Barbadori and Salviati. When enough of those low-degree families churn (due to the disparity threshold), the Medici then churn due to the activity threshold. Meanwhile, intermediate-degree families such as the Tornabuoni and Ridolfi continue sharing in the long term.

\subsubsection{Heatmap of Local Paradox vs Terminal Sharing Step}

\begin{figure}[h!]
  \centering
  \includegraphics[width=\linewidth]{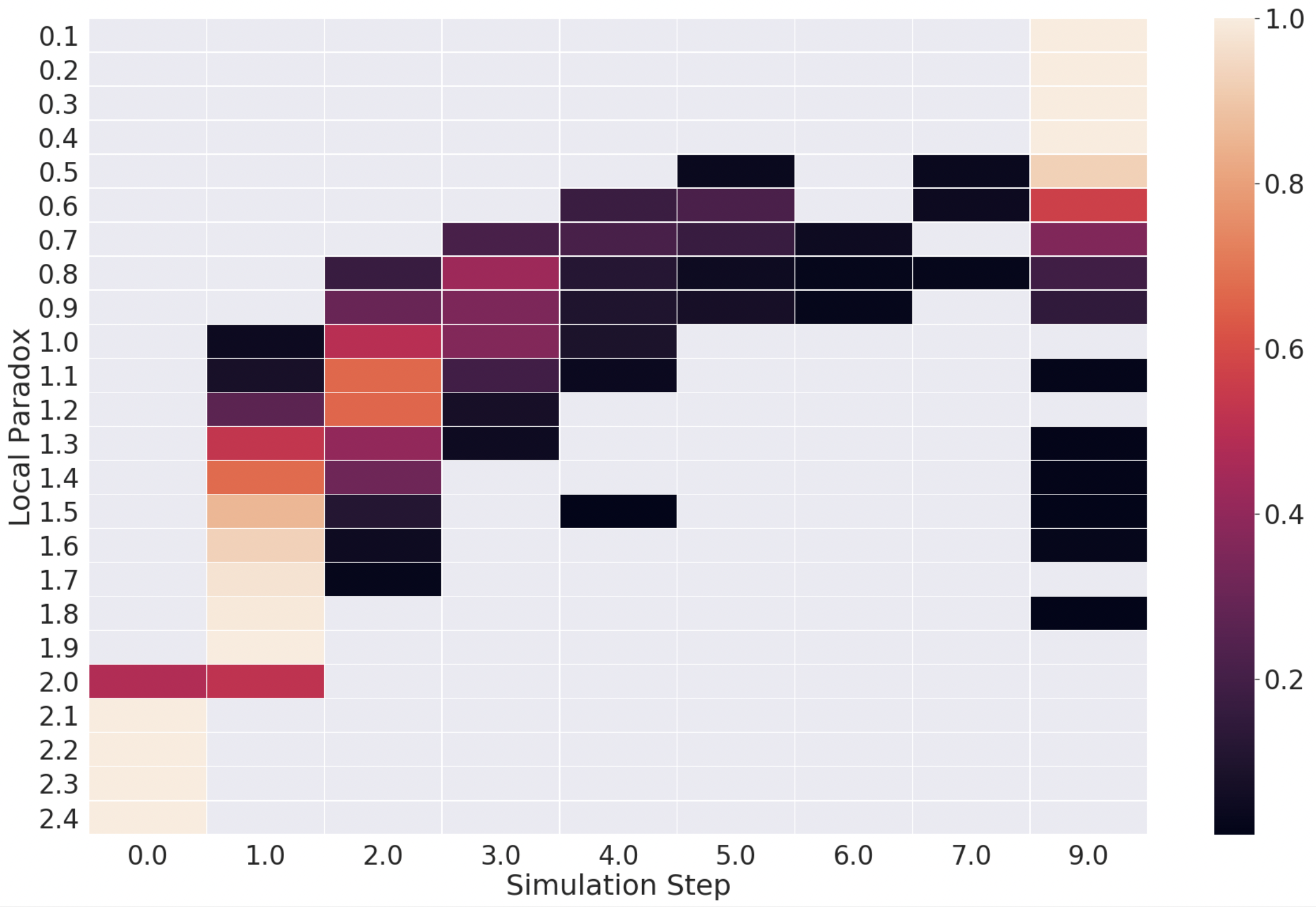}
  \caption{Initial Local Paradox values vs terminal sharing simulation step (grid color signifies fraction of people w y-axis paradox score who terminate at that step)}
  \label{heatmap_local_paradox}
\end{figure}

This heatmap presents visually the correlations between a person's local paradox and their terminal sharing step.\ref{heatmap_local_paradox}, with local paradox buckets on the y-axis and terminal simulation steps in the x-axis. Each grid value reflects the percentage of people with each horizontal local paradox whose last shared content is during the grid’s simulation step. We produce a heat map for the Barabási-Albert simulation with 3000 nodes and 140 edges, and a negative unit step function with a threshold of 2.0.

While correlations are high between how high the initial local paradox is and how late a person goes to 0 in the simulation, as observed earlier, however, there are many instances of deviation. For example, we see some people with a local paradox of 1.8 survive till the last step, whereas some people with a local paradox of 0.5 stop sharing after the 5th step.

\end{document}